\documentclass
[runinaddress,nobibnotes,nofootinbib,10pt,noshowpacs,noshowkeys,preprint,amsfonts,amssymb,tightenlines,final,titlepage,floatfix,preprintnumbers,twocolumn,twoside,balancelastpage]{revtex4}%
\usepackage{amssymb}
\usepackage{amsfonts}
\usepackage{amsmath}
\usepackage{graphicx}
\usepackage[mathscr]{euscript}
\usepackage{revsymb}
\usepackage{color,soul}

\setcitestyle{square}

\makeatletter
\newcommand*{\balancecolsandclearpage}{%
  \close@column@grid
  \clearpage
  \twocolumngrid
}
\makeatother
\usepackage{array}%
\providecommand{\U}[1]{\protect\rule{.1in}{.1in}}

\providecommand{\U}[1]{\protect\rule{.1in}{.1in}}

\begin{document}
\title{Metamaterials: $\boldsymbol{supra}$-classical dynamic homogenization}
\thanks{Dedicated with gratitude to the memory of Prof Yves C Angel}
\author{Mihai Caleap}
\author{Bruce W Drinkwater}
\affiliation{Faculty of Engineering, University of Bristol, BS8 1TR, United Kingdom}

\begin{abstract}
Metamaterials are artificial composite structures designed for controlling
waves or fields, and exhibit interaction phenomena that are unexpected on the
basis of their chemical constituents. These phenomena are encoded in effective
material parameters that can be electronic, magnetic, acoustic, or elastic,
and must adequately represent the wave interaction behaviour in the composite
within desired frequency ranges. In some cases -- for example, the low
frequency regime -- there exist various efficient ways by which effective
material parameters for wave propagation in metamaterials may be found.
However, the general problem of predicting frequency-dependent dynamic
effective constants has remained unsolved. Here, we obtain novel mathematical
expressions for the effective parameters of two-dimensional metamaterial
systems valid at higher frequencies and wavelengths than previously possible.
By way of an example, random configurations of cylindrical scatterers are
considered, in various physical contexts: sound waves in a compressible fluid,
anti-plane elastic waves, and electromagnetic waves. Our results point towards
a paradigm shift in our understanding of these effective properties, and
metamaterial designs with functionalities beyond the low-frequency regime are
now open for innovation.

\end{abstract}
\maketitle

\section*{Introduction}

Metamaterial research in the past decade offered an entirely new route to
further enhance our capability to engineer material properties at will. Here,
metamaterials are artificially fabricated structures (often periodic,
\textit{i.e.}, crystalline) which are designed so that they exhibit wave
properties not observed with common materials, \textit{e.g.}, they can, in
theory, bend electromagnetic \cite{Leonhardt2006}, acoustic \cite{Li2009}, and
even surface gravity waves \cite{Porter2014} so as to achieve sub-wavelength
focusing \cite{Pendry2000}, create cloaks \cite{Pendry2006,Milton2006}, and
attain shielding \cite{Feng2008a}. Other unexpected properties include
artificial magnetism \cite{Pendry1999}, negative permeability \cite{Smith2000}%
, negative refraction index \cite{Shelby2001}, and hyperbolic dispersion
\cite{Liu2007}, to name a few. Such materials have allowed us to gain
unprecedented control over a range of electromagnetic/optical and acoustic
wave phenomena. In many ways metamaterials parallel the development of
photonic and phononic crystals (optical and acoustic analogues of
semiconductors) which also rely on small-scale structures for their
properties. However, the major difference lies in the sub-wavelength nature of
metamaterial structure. This enables us to summarize their properties in terms
of permittivity and permeability $\left(  \varepsilon,\mu\right)  $ for
electromagnetic waves, or bulk modulus and mass density $\left(  \kappa
,\rho\right)  $ for acoustic waves, just as we would for any naturally
occurring material. This is an enormous simplification for the design process,
and research is now focusing on the realization of a new generation of
metadevices \cite{Zheludev2012}\ with novel and useful functionalities
achieved by the structuring of functional matter on the sub-wavelength scale.
Novel devices such as superlens \cite{Zhang2008}, hyperlens \cite{Lu2012},
invisibility cloaks \cite{Landy2013,Zigoneanu2014}, and plasmonic waveguides
\cite{Barnes2003}\ have been fabricated and tested in the past few years. The
technology behind such metadevices is fairly well established in the
\textit{low-frequency} regime where inclusions have sizes much smaller than
the wavelength of operation. At these relatively low frequencies this is
commonly obtained by assuming only monopole and/or dipole interactions,
\textit{e.g.}, by utilizing conducting materials shaped as dipoles
\cite{Pendry1996}\ and split-ring resonators \cite{Pendry1999a}. The existence
of resonances poses a considerable challenge to classical effective medium
theories. This is because their basic principle is to minimize the scattering
in the quasi-static limit, while the local resonances occur most often at
longer wavelengths.

Here, following these concepts, we develop and analyse a \textit{supra}%
\footnote{This word comes from Latin and means \textit{above} or
\textit{beyond the limits of}.}-classical dynamic model of metamaterial
response. There is an abundance of miscellaneous effective medium theories
\cite{Li2004,Wu2006,Wu2007,Torrent2006,Mei2006,Torrent2011,Wu2012}, some quite
recent \cite{Torrent2014,Zhang2015,Savoia2015a}; many of these works claim to
be valid not only in the quasi-static limit but also at finite frequencies
\textit{beyond} the long-wavelength limit: a situation that happens when the
wavelength $\Lambda$ is long in the host medium, while the wavelength in the
particles, $\Lambda_{0}$, can be small. (This is in contrast to the
quasi-static limit where both $\Lambda$ and $\Lambda_{0}$ should be much
larger than the size of the particles.) Such extension to finite frequencies
is sometimes denoted as the dynamic effective medium theory. However, even
this \textit{dynamic} approximation relies exclusively on the monopolar and
dipolar response of the scattering objects, which implicitly assumes long
wavelengths. In this paper, this restriction is relaxed and the full effect of
the ensemble of particles that constitute the effective medium is included, as
higher diffraction orders are encompassed. This will allow the design of new
metadevices working over a wider wavelength range. We shall illustrate this by
solving a simple scalar problem in two dimensions, having applications not
only in electromagnetics but also in acoustics and elasticity. The
similarities between the equations of acoustics, elasticity and
electromagnetics allow us to use some of the same techniques to solve problems
in these seemingly disparate fields.

\section*{View on classical homogenization}

The theoretical approach to the field of metamaterials is provided through
dynamic homogenization techniques which relate the microstructure of a
composite to the frequency dependence of its effective properties. The
majority of research interest in the area of metamaterials is restricted to
\textit{periodic} microstructures \cite{Yablonovitch1987a,Martinez-Sala1995}
(as the arrangement of molecules according to solid-state physics) which admit
Bloch (or Floquet) waves\ as solutions and many different numerical algorithms
have been developed (see, \textit{e.g.}, \cite{Ho1990,Goffaux2003}) for
calculating the dispersive properties of these waves. A popular route to
determining these parameters is by the use of retrieval methods
\cite{Smith2002,Fokin2007} where the assumption is that local effective
properties may be used to define periodic composites. The retrieval method
leads to the refractive index $n$ and the wave impedance $\mathcal{Z}$, which
defines the reflectivity of a semi-infinite slab. However, while simple in
principle, such retrieval methods are limited to ordered arrays and often
produce ambiguous results due to oversimplified initial assumptions of the
bulk model \cite{Woodley2010}.

Certainly engineers like structures and designs that follow some type of
order. However, materials may be also amorphous and isotropic, and natural
materials on the macroscopic level are quite often \textit{random} in essence.
It may well be that a random placement of complex particles would be enough to
produce emergent properties in the overall wave response and therefore give us
a sample of metamaterial \cite{Caleap2012}. The effective behaviour of
metamaterials whose microstructure is random depends strongly on the governing
statistics of the random distribution. Effective properties may be determined
by using the self-consistent \textit{effective medium} methods for which a
substantial body of literature may be found. Although variants exist, these
methods often consider the scattering problem of a coated particle embedded in
a matrix which has the properties of the effective media. These properties are
then determined by requiring the vanishing of the effective forward-scattering
amplitude $f_{0}^{eff}=f_{\theta=0}^{eff}$\ and as such are formally
restricted to the low-frequency and long-wavelength ranges. For examples where
this method has been applied to electromagnetic, acoustic and elastic waves,
see \cite{Li2004,Wu2006,Wu2007,Jin2009a,Jin2012,Zhang2015}. Although the above
self-consistent condition ($f_{0}^{eff}=0$) is physically sufficient to
describe the effective medium, two effective properties, \textit{i.e.
}$\left(  \varepsilon_{eff},\mu_{eff}\right)  $, cannot be determined
`simultaneously and uniquely' from the single condition. A supplementary
condition is needed; this prevents the application of effective medium methods
to finding dynamic effective properties. Note however that the above condition
is sufficient for wave propagation in a metamaterial in which a single
material constant is involved, \textit{e.g.} in dielectric media. Another
deficiency of many current enhancements of the effective medium methods is
their failure to describe the influence of the spatial distribution of
particles on the effective constitutive parameters. Such a description is
possible in the framework of a self-consistent scheme called the
\textit{effective field }method \cite{Caleap2012}\ and our work is within the
framework of this scheme. One of the principal results of the effective field
approach was an adequate definition of the coherent wave and a proof that it
obeys a wave equation, \textit{i.e.}, a proof that, under certain conditions,
a random distribution of scatterers can, for this purpose, be represented by
an effective medium \cite{Foldy1945}. Most calculations proceed by assuming
the existence of such an effective medium equation.

The subject of the present work is the macroscopic dynamic behaviour of the
above composite medium, \textit{i.e.}, random distribution of particles. More
precisely, we shall describe a heuristic scheme for evaluating the effective
properties of metamaterials. The approach is based on the idea that a certain
effective field acts on each particle, as a consequence of the presence of the
other particles; hence, the name \textit{effective field} method\textit{.} The
framework we develop is based on the Fikioris-Waterman \cite{Fikioris1964,
Fikioris2013} and Waterman-Pedersen \cite{Waterman1986}\ formalism to evaluate
the coherent wave motion on both sides of a semi-infinite array of
particles.\footnote{The later reference, is the earliest work to our knowledge
to predict explicit relations for the effective bulk parameters $\left(
\varepsilon_{eff},\mu_{eff}\right)  $ in the dynamic range. The authors have
also predicted negative frequency-dependent $\mu_{eff}$ at single-particle
resonances although the plots only displayed the positive values. In fact,
they only noted that \textquotedblleft\textit{the effective parameters vanish
or diverge at certain frequencies}\textquotedblright\ without further comment,
which suggests that the results were considered curious at that time.
Currently it is common to have negative effective parameters, and much
research on metamaterials is focused on this area.} More specifically, we
consider an averaged wave motion, where all possible configurations of
particles are weighted by appropriate pair-correlation functions. In contrast
to the effective medium methods, we derive a fully dynamic model for the
effective constitutive parameters, which retains all the relevant information
(particle geometry and physical parameters)\ provided by an expanded multipole
solution. As a result, the theory discussed in the following is more complete
and potentially more useful than previous approaches to derive effective
material parameters.

\section*{Results}

Here, we consider two specific polarizations in electromagnetism, transverse
electric (\textsc{te}) and transverse magnetic (\textsc{tm}). In addition,
parallel to the electromagnetic example is the mathematically identical case
of acoustics and anti-plane elasticity.

We then consider these as two dimensional problems. Indeed, exploiting the
physics common to many types of wave propagation, the idea of metamaterials
has been implemented successfully for acoustic and elastic waves. Many of the
conclusions drawn from photonics research directly apply to acoustic waves and
acoustic metamaterials due to the essential similarity of the governing
equations in the two cases. Realizing analogous results for elastic
metamaterials is complicated by the fact that the governing equation for
elasticity admits both longitudinal and shear wave solutions which are capable
of exchanging energy between each other. However, anti-plane elasticity is a
special state of deformation with just a single non-zero displacement field,
similar to transverse electromagnetics. The governing equation common to
electromagnetics, acoustics, and linear anti-plane elasticity is detailed in
Appendix A.

Effective constitutive parameters depend on many factors including the
intrinsic properties of the particles and the host matrix, their shape and
topology. The latter determines how the particles are distributed in the
matrix. The system considered in our study is composed of two isotropic
phases: cylindrical particles of arbitrary shape randomly distributed in a
host medium with propagation constant $k=\omega\sqrt{\mathsf{md}}$ for some
(possibly complex) parameters ($\mathsf{m},\mathsf{d}$)\ of the medium.
Depending on the application, these material parameters could be,
\textit{e.g.}, compliance ($1/G$) and density ($\rho$) for shear horizontal
polarized elastic waves or permittivity ($\varepsilon$)\ and permeability
($\mu$) in electromagnetism; a number of useful relationships among these
parameters are summarized in Table I.

\begin{table}[ptb]
\caption{Relationships among electromagnetic, acoustic and elastic material
parameters.\medskip}%
\centering{\small \addtolength{\tabcolsep}{4pt}
\begin{tabular}
[c]{@{\vrule height 11.5pt depth 4pt width 0pt}crccc}\cline{2-5}\cline{2-5}
&
\multicolumn{2}{r}{Electromagnetics\tablenote{Observe that the permittivity and permeability for a specific polarization can be related to a pair of acoustic and elastic constants. For instance, $\left( \varepsilon ^{\text{\textsc{TE}}},\mu ^{\text{\textsc{TE}}}\right) \leftrightarrow (\rho ^{\text{\textsc{SH}}},\rho ^{\text{\textsc{P}}})\leftrightarrow (1/\kappa ,1/G)$.}}
& Acoustics & Elasticity\\\hline\hline
{\textsf{m}\quad} & {$\varepsilon^{\text{\textsc{TM}}}$ } & {$\mu
^{\text{\textsc{TE}}}$ } & $\rho^{\text{\textsc{P}}}${\ } & {$1/G$ }\\
{\textsf{d}\quad} & {$\mu^{\text{\textsc{TM}}}$ } & {$\varepsilon
^{\text{\textsc{TE}}}$ } & {$1/\kappa$ } & {$\rho^{\text{\textsc{SH}}}$
}\\\hline
\end{tabular}
}\end{table}

In Appendix B, we briefly review the effective field method. Subject to the
quasi-crystalline approximation, two equations are obtained for which the
effective wavenumber $\mathcal{K}$ of some coherent wave motion (either
electromagnetic, acoustic, or elastic) and the effective impedance
$\mathcal{Z}$, are given in implicit form. Note that, whereas the dispersion
relation for $\mathcal{K}$ is polarization-independent, the effective
impedance depends on the type of the incident wave. These equations are the
starting point of all further developments. Observe that the particles have a
size distribution and their relative positions are described by an arbitrary
\textit{cross-pair} distribution function $g_{ij}$. Also, the size
distribution is represented by $\eta_{j}=\eta(a_{j})$; here $a_{j}$ is the
radius of the circular surface\ circumscribing a particle, and $\eta$ is the
number of particles per unit area.

Without loss of generality, we next assume the particles are identical and
have equal sizes $a_{j}\equiv a$. Here, we refer only to the final explicit
solutions for the effective parameters $\mathsf{m}_{eff}$ and $\mathsf{d}%
_{eff}$, which are expressed elegantly as%

\begin{equation}
\frac{\mathsf{m}_{eff}}{\mathsf{m}}\simeq1+\widetilde{\mathsf{m}}_{1}%
\frac{\epsilon}{2k^{2}}+\widetilde{\mathsf{m}}_{2}\frac{\epsilon^{2}}{2k^{2}%
}+{\mathcal{O}}\left(  \epsilon^{3}\right)  , \label{m}%
\end{equation}%
\begin{equation}
\frac{\mathsf{d}_{eff}}{\mathsf{d}}\simeq1+\widetilde{\mathsf{d}}_{1}%
\frac{\epsilon}{2k^{2}}+\widetilde{\mathsf{d}}_{2}\frac{\epsilon^{2}}{2k^{2}%
}+{\mathcal{O}}\left(  \epsilon^{3}\right)  , \label{d}%
\end{equation}
where $\epsilon=4\pi\eta$. The scalar coefficients $(\widetilde{\mathsf{m}%
}_{1},\widetilde{\mathsf{m}}_{2})$ and $(\widetilde{\mathsf{d}}_{1}%
,\widetilde{\mathsf{d}}_{2})$ are given in matrix notation by%

\begin{subequations}
\label{meeff}%
\begin{align}
\widetilde{\mathsf{m}}_{1}  &  ={{\mathbf{e}}}^{t}\mathbf{Qe}-{{\mathbf{e}}%
}^{t}{\mathbf{J}}\mathbf{Qe,}\\
\widetilde{\mathsf{m}}_{2}  &  ={{\mathbf{e}}}^{t}%
\mathbf{Q\widetilde{{\mathbf{R}}}Qe}-{{\mathbf{e}}}^{t}{\mathbf{JQ}%
}\widetilde{{\mathbf{R}}}{\mathbf{Q}}\mathbf{e}\nonumber\\
&  -\frac{1}{4k^{2}}\left[  {\left(  {{\mathbf{e}}}^{t}\mathbf{Qe}\right)
}^{2}-\left(  {{\mathbf{e}}}^{t}{\mathbf{J}}\mathbf{Qe}\right)  ^{2}\right]  ,
\end{align}
and%

\end{subequations}
\begin{subequations}
\label{deeff}%
\begin{align}
\widetilde{\mathsf{d}}_{1}  &  ={{\mathbf{e}}}^{t}\mathbf{Qe}+{{\mathbf{e}}%
}^{t}{\mathbf{J}}\mathbf{Qe,}\\
\widetilde{\mathsf{d}}_{2}  &  ={{\mathbf{e}}}^{t}{\mathbf{Q}}%
\widetilde{{\mathbf{R}}}{\mathbf{Q}}\mathbf{e}+{{\mathbf{e}}}^{t}{\mathbf{J}%
}\mathbf{Q\widetilde{{\mathbf{R}}}Qe}\nonumber\\
&  -\frac{1}{4k^{2}}\left[  {\left(  {{\mathbf{e}}}^{t}\mathbf{Qe}\right)
}^{2}-\left(  {{\mathbf{e}}}^{t}{\mathbf{J}}\mathbf{Qe}\right)  ^{2}\right]  .
\end{align}

One can easily check that these equations are compatible when $\mathsf{m}%
_{eff}\mathsf{d}_{eff}\mathcal{K}^{2}=\omega^{2}$. Incidentally, we obtain
$\mathcal{K}^{2}\simeq k^{2}+\epsilon{{\mathbf{e}}}^{t}\mathbf{Qe}%
+\epsilon^{2}{{\mathbf{e}}}^{t}{\mathbf{Q}}\widetilde{{\mathbf{R}}}%
{\mathbf{Q}}\mathbf{e}+{\mathcal{O}}\left(  \epsilon^{3}\right)  $, which is,
as expected, the second order expansion in $\epsilon$ of the implicit
wavenumber equation \eqref{keff}. Note that all notations appearing in Eqs.
\eqref{meeff} and \eqref{deeff} are introduced in the Appendix.

Results in classical multiple scattering theories are usually defined in terms
of the \textit{angular shape function} $f_{\theta}$ for scattering of a plane
wave by a single particle. It is useful to render yet another form of the
coefficients \eqref{meeff} and \eqref{deeff} in terms of $f_{\theta}$. This is
done by considering the line-like approximation: in addition to $\epsilon
a^{2}\ll1$, we also require $kb\ll1$. To render the results more tractable,
the spatial distribution of particles is assumed to be isotropic and
homogeneous, for which $g_{ij}\left(  r\right)  \equiv g\left(  r\right)
=\mathrm{H}(r-b)$. This describes a non-overlapping condition; here, $g$
denotes a \textit{pair}-correlation function, $\mathrm{H}$ is the Heaviside
unit function, and $b=2a$ is the diameter of the particles. Retaining only the
leading order term in $kb$ of the multiple scattering matrix
$\mathbf{\widetilde{{\mathbf{R}}}}$, and using the definition \eqref{ftheta}
for $f_{\theta}$, we obtain
\end{subequations}
\begin{equation}
{{\mathbf{e}}}^{t}{\mathbf{Q}}\widetilde{{\mathbf{R}}}\mathbf{Qe}\cong%
-\frac{\mathrm{1}}{4k^{2}}%
\mathscr{H}%
_{0}\text{ and }{{\mathbf{e}}}^{t}\mathbf{JQ\widetilde{{\mathbf{R}}}Qe}%
\cong-\frac{\mathrm{1}}{4k^{2}}%
\mathscr{H}%
_{\pi}, \label{ho}%
\end{equation}
with%

\begin{equation}%
\mathscr{H}%
_{\alpha}=\frac{2}{\pi}\int_{0}^{\pi}\mathrm{d}\theta\cot(\theta
/2){\tfrac{\mathrm{d}}{\mathrm{d}\theta}}%
\mathscr{G}%
_{\alpha}^{\theta}\text{ and }%
\mathscr{G}%
_{\alpha}^{\theta}=f_{\theta}f_{\alpha-\theta}. \label{H}%
\end{equation}

By means of these approximations, we can infer the following closed-form
constitutive relations%

\begin{align}
\frac{\mathsf{m}_{eff}}{\mathsf{m}}  &  \simeq1+\frac{\epsilon}{2k^{2}}\left(
f_{0}-f_{\pi}\right) \nonumber\\
&  +\frac{\epsilon^{2}}{8k^{4}}\left[  \left(
\mathscr{G}%
_{0}^{\pi}-%
\mathscr{G}%
_{0}^{0}\right)  -\left(
\mathscr{H}%
_{0}-%
\mathscr{H}%
_{\pi}\right)  \right]  , \label{m0}%
\end{align}%
\begin{align}
\frac{\mathsf{d}_{eff}}{\mathsf{d}}  &  \simeq1+\frac{\epsilon}{2k^{2}}\left(
f_{0}+f_{\pi}\right) \nonumber\\
&  +\frac{\epsilon^{2}}{8k^{4}}\left[  \left(
\mathscr{G}%
_{0}^{\pi}-%
\mathscr{G}%
_{0}^{0}\right)  -\left(
\mathscr{H}%
_{0}+%
\mathscr{H}%
_{\pi}\right)  \right]  . \label{d0}%
\end{align}

Apart from their dependence on $k$ and $\epsilon$ (or $\eta$), the effective
dynamic parameters ($\mathsf{m}_{eff},\mathsf{d}_{eff}$) given by Eqs.
\eqref{m0} and \eqref{d0} are all completely determined when the angular shape
function $f_{\theta}$\ for an isolated particle is known. If this scattering
amplitude can be determined either analytically, numerically, or
experimentally, then the effective medium equivalent to the artificial
composite is \textit{fully} described.

It is noteworthy that if one wants to study the behaviour of effective
parameters at high concentrations (where such expansions may not be valid) the
general implicit equations detailed in Appendix B should be used and/or more
accurate pair correlation functions should be considered. Neither incident
wave nor boundary conditions have entered yet in the above description.
Consequently, the results admit several solutions corresponding to different
polarization states. In Ref. \cite{Supp} (section\ S1), the expansions
\eqref{m} and \eqref{d} [or \eqref{m0} and \eqref{d0}] are specialized to
electromagnetic, acoustic, and elastic scattering for long wavelengths
($a\ll\Lambda$). This provides an additional check on the correctness of the
results obtained in this paper. A further check on the consistency of our
method is provided in Ref. \cite{Supp} (section S2). It is shown that the
quasi-crystalline approximation is \textit{self-consistent} and identical to
coherent potential approximation \cite{Milton1985}\ at least to second order
in concentration, provided the effective parameters are identified as those
derived in this section.

\section*{Discussion}

While the limiting cases considered in Ref. \cite{Supp} (section S1) perform a
check of the theory we have presented, they neglect some important features of
the effective field method. Therefore, we address this problem numerically in
order to illustrate the dynamic behaviour of the effective parameters. In the
following, the effective parameters $\left(  \mathsf{m}_{eff},\mathsf{d}%
_{eff}\right)  $ are calculated by using Eqs. \eqref{m} and \eqref{d},
together with the Percus-Yevick pair-correlation function for hard disks
\cite{Adda-Bedia2008}.

\subsubsection*{Example illustration}

We consider a fibre bundle (or circular cluster of dielectric fibres with
$\mu_{0}=1$) of effective radius $r_{eff}$ in vacuum. A plane electromagnetic
wave is incident on the fibre bundle. A sketch is shown in Figure 1. There are
$68$ circular fibres each of radius $a$ randomly distributed in the cluster
and their volume fraction is $10.88\%$. The refractive index of the fibres is
$\emph{n}_{0}$($=\sqrt{\varepsilon_{0}}$)$=1.33+0.01\mathrm{i.}$ Exact
multiple scattering simulations\footnote{The analytical solution to Maxwell
equations for scattering by multiple parallel cylinders has been described,
\textit{e.g.}, in Ref. \cite{Felbacq1994a}.} are compared with the effective
medium model (\textit{i.e.} equivalent homogeneous magneto-dielectric
inclusion\footnote{Note that, as expected, our results also predict an
effective magnetic permeability $\mu_{eff}$ at finite frequencies (different
from that in vacuum) in a system in which both the matrix and the particles
are non-magnetic.} with effective parameters $\left(  \varepsilon_{eff}%
,\mu_{eff}\right)  $). The multiple scattering results are averaged over
different realizations of the fibres locations. With $500$ realizations, the
maximum error between the numerical model and the effective medium results is
less that $0.5\%$, for the two cases illustrated.

\begin{figure}[ptb]
\centerline{\includegraphics[width=.5\textwidth]{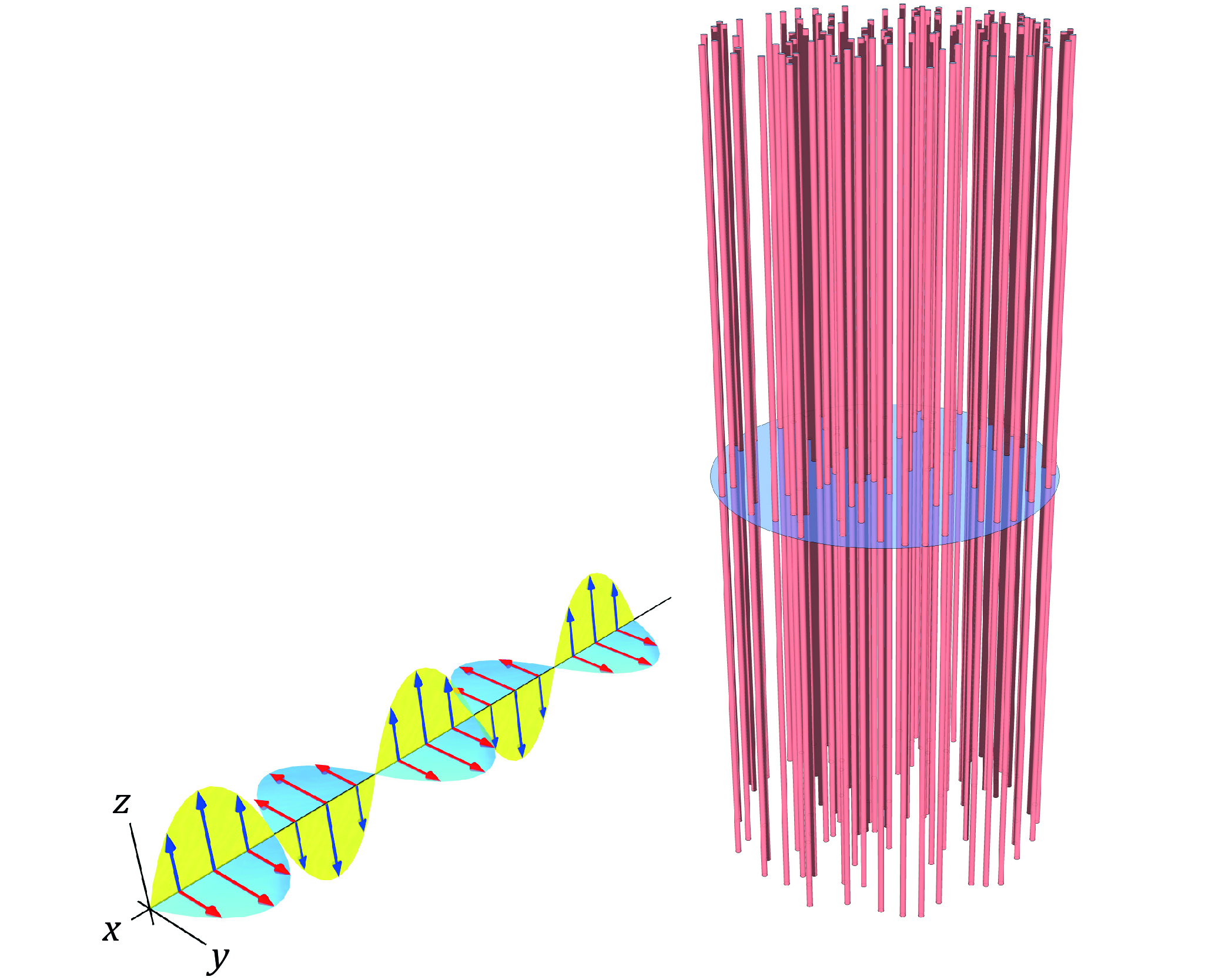}}\centerline{}\caption{Sketch of the
scattering of an electromagnetic wave by a circular cluster of cylindrical
particles.}%
\end{figure}

Figures 2 and 3 show the spatial maps of the near-field electric field
amplitude $\left\vert E_{z}\right\vert $ for two different incident
wavelengths, $\Lambda=2r_{eff}/5=10a$ and $\Lambda=4r_{eff}/25=4a$,
respectively. Figure 2 illustrates the response of the coherent wave regarding
the topology of the fibre-bundle. As expected, the waves are insensitive to
the relative locations of the fibres for long wavelengths. This is not the
same for shorter wavelengths. A comparison of the results in Figure 3
indicates the agreement is excellent even for the high frequency case
considered ($\Lambda=4a$). It is particularly encouraging that the agreement
is excellent even inside the circular cluster. Observe that the a regular
arrangement of fibres produces a result that is closer to the effective
cluster for long wavelengths, than is the result obtained with a random
realization of the fibres locations. We should note that although the
comparison in Fig. 3 is excellent, it may not always be so for other
geometries of the fibre bundle. In a final section we detail various
limitations and assumptions of our model and discuss other similar problems
obtained previously.

\begin{figure}[ptb]
\centerline{\includegraphics[width=.5\textwidth]{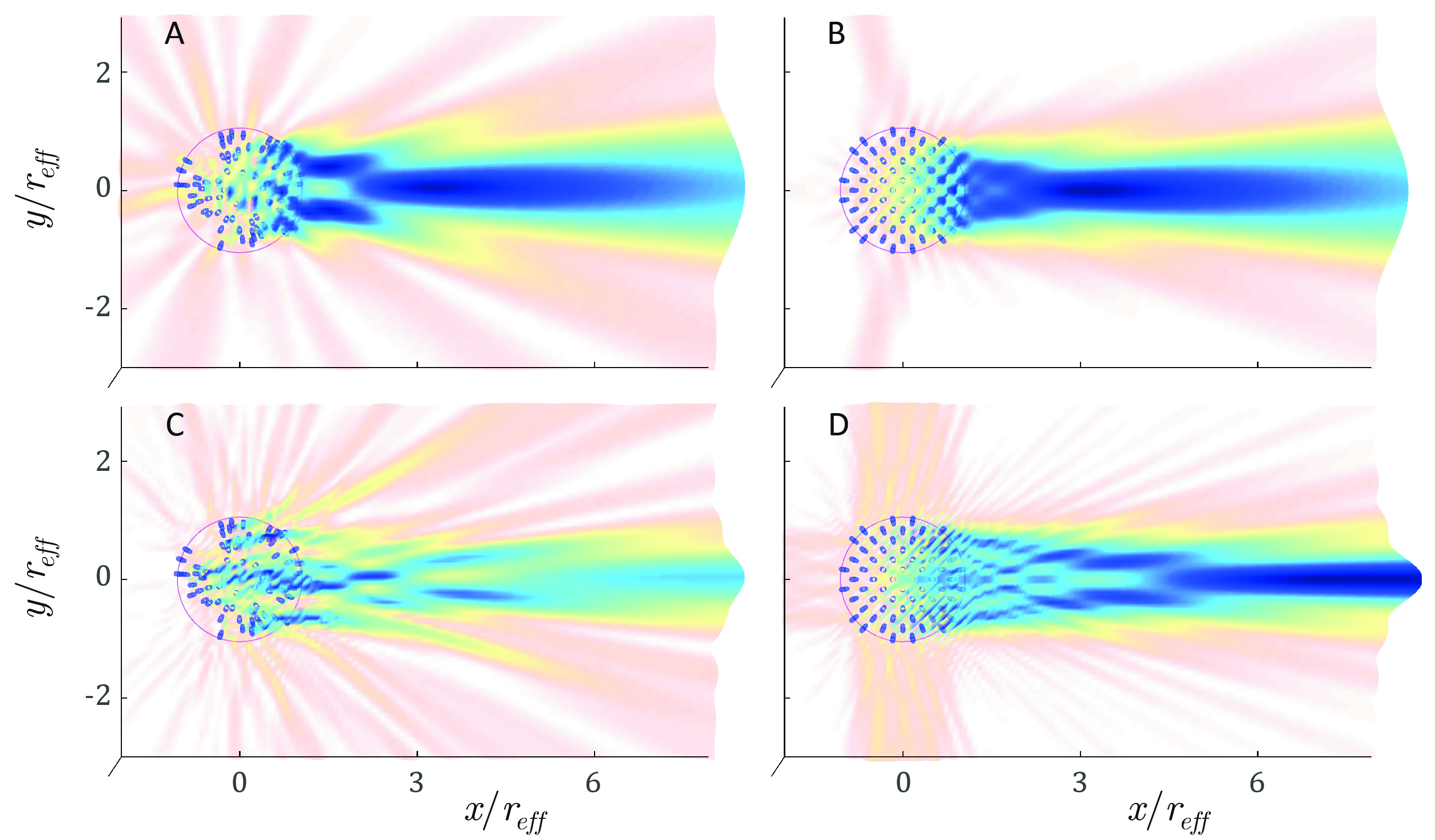}}\centerline{}\caption{Spatial
maps of near-field electric field $\left\vert E_{z}\right\vert $ as a
\textsc{te} wave is incident from the left. Exact multiple scattering
simulations for a single realization of fibres locations: left panels (A, C)
-- random array; right panels (B, D) -- regular array. Top panels (A, B) --
effective radius $r_{eff}$ of the cluster is such that $\Lambda=2r_{eff}%
/5=10a$; bottom panels (C, D) -- $\Lambda=4r_{eff}/25=4a$.}%
\label{Fig2}%
\end{figure}

\begin{figure}[ptb]
\centerline{\includegraphics[width=.5\textwidth]{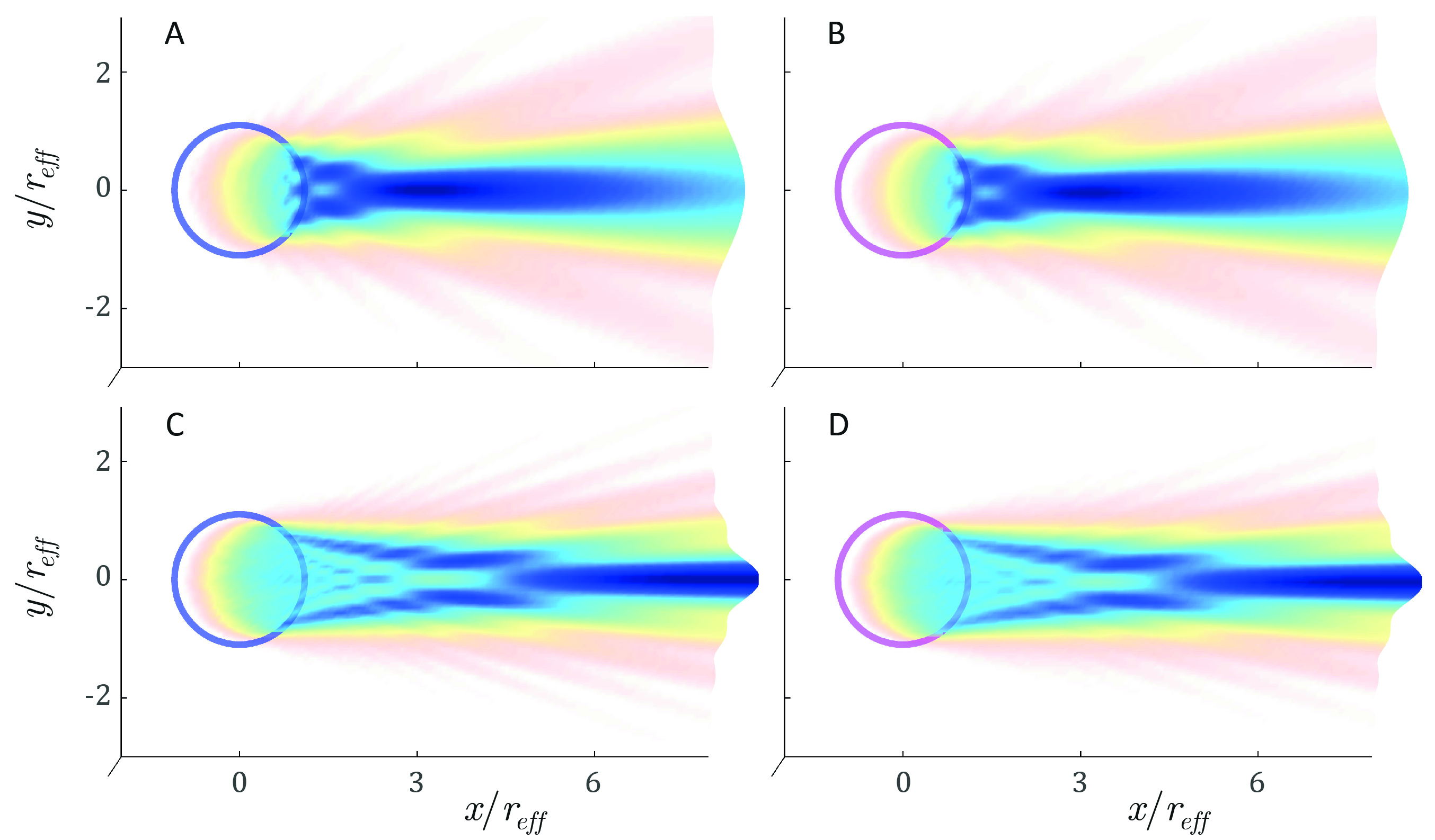}}\centerline{}\caption{{Spatial
maps of near-field electric field }$\left\vert {E_{z}}\right\vert ${ as a
\textsc{te} wave is incident from the left. Left panels (A, C) -- average over
500 different realizations (of exact multiple scattering simulations); right
panels (B, D) -- equivalent homogeneous inclusion (single scattering result)
with dynamic effective parameters. Top panels (A, B) -- effective radius
}$r_{eff}${ of the cluster is such that $\Lambda=2r_{eff}/5=10a$; bottom
panels (C, D) -- $\Lambda=4r_{eff}/25=4a$.}}%
\label{Fig3}%
\end{figure}

\subsubsection*{Anisotropic metamaterials}

It is of considerable interest to discuss the possibility of realizing
anisotropic metamaterials, that is, the\ material parameters are not scalars
but tensors, with their principle components taking different values.
Different from the anisotropy property of the material itself, we shall
examine anisotropy originating from geometric asymmetry and consider a random
array of elliptic cylinders of material parameters ($\mathsf{m}_{0}%
,\mathsf{d}_{0}$). The $x$- and $y$-axes are set in the directions of the
semi-minor and semi-major axes of the elliptic cylinders, with respective
radii $a_{x}$ and $a_{y}$. Due to the geometric arrangement of the elliptic
cylinders and the symmetry of the scattered fields, the $x$- and
$y$-directions can therefore be seen as effective principal directions.
Proceeding essentially as detailed in Ref. \cite{Supp} (section S1), we
obtain, in the quasi-static limit%

\begin{align}
\frac{\mathsf{m}_{eff,x}}{\mathsf{m}}  &  \simeq1+2\phi\mathcal{M}%
_{x}\mathcal{+}2\phi^{2}\mathcal{M}_{x}^{2},\label{mx}\\
\frac{\mathsf{m}_{eff,y}}{\mathsf{m}}  &  \simeq1+2\phi\mathcal{M}%
_{y}\mathcal{+}2\phi^{2}\mathcal{M}_{y}^{2},\\
\frac{\mathsf{d}_{eff}}{\mathsf{d}}  &  \simeq1+\phi\mathcal{D}, \label{dx}%
\end{align}
where $\phi=\pi\eta a_{x}a_{y}$ is the volume fraction of the elliptical
cylinders. The coefficients $\mathcal{D}$ and ($\mathcal{M}_{x},\mathcal{M}%
{_{y}}$) are given by%

\begin{equation}
\mathcal{D}=\frac{\mathsf{d}_{0}}{\mathsf{d}}-1\text{ and }\binom
{\mathcal{M}{_{x}}}{\mathcal{M}{_{y}}}=\frac{1}{2}\frac{\left(  \mathsf{m}%
_{0}-\mathsf{m}\right)  \left(  a_{x}+a_{y}\right)  }{\mathsf{m}_{0}%
\binom{a_{x}}{ay}+\mathsf{m}\binom{a_{y}}{a_{x}}}.
\end{equation}

Observe that if $a_{x}=a_{y}$ ($=a$, \textit{i.e.}, circular cross section)
then $\mathcal{M}{_{x}=}\mathcal{M}{_{y}}$ (${=}\mathcal{M}$, see Ref.
\cite{Supp}, section S1), and $\mathsf{m}_{eff,x}=\mathsf{m}_{eff,y}$. The
results of Eqs. \eqref{mx}-\eqref{dx} show that only the effective property
$\overleftrightarrow{\mathsf{m}}_{eff}$ is a tensor with principal components
$\mathsf{m}_{eff,x}$ and $\mathsf{m}_{eff,y}$, whereas $\mathsf{d}_{eff}$ is a
scalar. This is consistent with results obtained recently in Ref.
\cite{Zhang2015}, for electromagnetic waves in the quasi-static limit.
However, these results should be consumed with prudence. We show that, in
general, both $\overleftrightarrow{\mathsf{m}}_{eff}$ and
$\overleftrightarrow{\mathsf{d}}_{eff}$ are tensors for arbitrary frequency
and wavelength. To see this more clearly, let us consider the scattering of a
\textsc{tm} wave by perfect electric conductive elliptic cylinders in
vacuum\footnote{This is equivalent to solving the Neumann boundary\ condition.
In anti-plane elasticity, this condition corresponds to a cylindrical cavity
with stress-free surface.}. From Table I, we infer that
($\mathsf{m}$,$\mathsf{d}$) corresponds to ($\varepsilon,\mu$), for
\textsc{tm} waves; appropriate identifications the resulting effective medium
are implied. Figure 4 shows the effective permittivity
$\overleftrightarrow{\varepsilon}_{eff}$ and permeability
$\overleftrightarrow{\mu}_{eff}$ tensors. Only the real part of these
parameters is presented for brevity. The volume fraction $\phi$ is fixed and
equal to $6\pi\%$. It should be noted that the actual concentration $\phi
=\pi\eta a_{x}a_{y}$ cannot exceed $a_{x}/a_{y}$ in order to be consistent
with our model, so that $\phi=\pi\eta a^{2}\leq1$ when $a_{x}=a_{y}$ ($=a$).
The figure is intended to illustrate the variations of
$\overleftrightarrow{\varepsilon}_{eff}/\varepsilon$ and
$\overleftrightarrow{\mu}_{eff}/\mu$ as the wavelength $\Lambda/\widetilde{a}$
varies on the horizontal axis, for several aspect ratios $a_{x}/a_{y}$; here,
$\widetilde{a}=\sqrt{a_{x}a_{y}}$ is the geometric mean of the semi-minor and
semi-major axes, $a_{x}$ and $a_{y}$. Observe that in the quasi-static limit,
for $\Lambda>10\widetilde{a}$, where currently available model will be
adequate, the principal components of $\overleftrightarrow{\mu}_{eff}$ are
visibly equal, \textit{i.e.}, $\mu_{eff,x}\simeq\mu_{eff,y}$, regardless of
the ratio $a_{x}/a_{y}$. This is as expected, given Eq. \eqref{dx}. It is
interesting that for shorter wavelengths ($\Lambda<10\widetilde{a}$),
$\mu_{eff,x}$ and $\mu_{eff,y}$ become increasingly distinct as the ratio
$a_{x}/a_{y}$ decreases from $1$ to $0.5$, an effect not predicted by the
existing literature. This suggests a new route to the design of metamaterials
with controllable anisotropic effective properties.

\begin{figure}[ptb]
\centerline{\includegraphics[width=.5\textwidth]{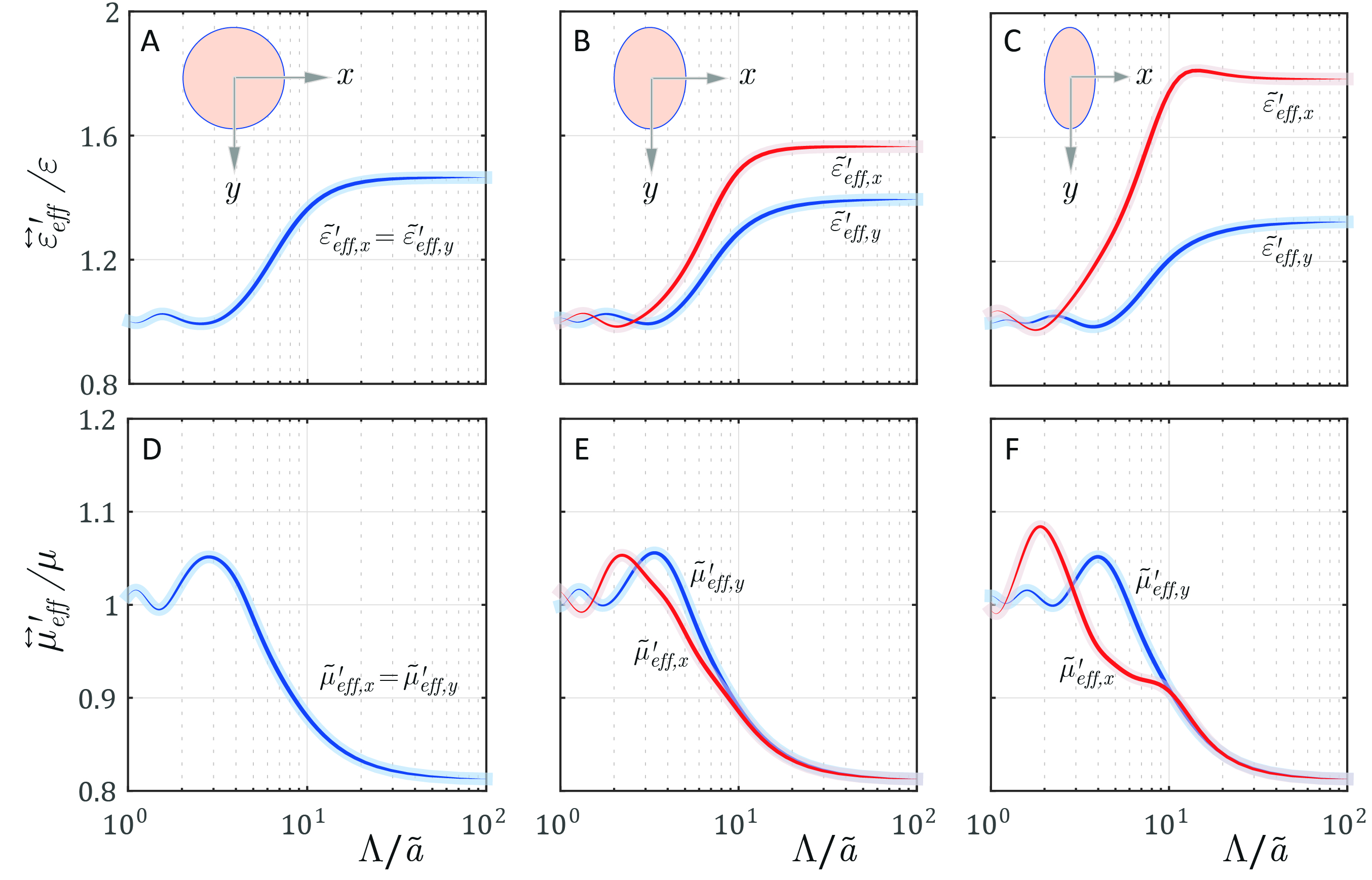}}\centerline{}\caption{{Effective
permittivity $\protect\overleftrightarrow{\varepsilon}_{eff}/\varepsilon$
($=\protect\widetilde{\varepsilon}_{eff}^{\prime}+\mathrm{i}%
\protect\widetilde{\varepsilon}_{eff}^{\prime\prime}$) (top panels, A-C) and
permeability $\protect\overleftrightarrow{\mu}_{eff}/\mu$
($=\protect\widetilde{\mu}_{eff}^{\prime}+\mathrm{i}\protect\widetilde{\mu
}_{eff}^{\prime\prime}$) (bottom panels, D-F) \textit{versus} the wavelength
for a \textsc{tm}-polarized field incident upon a random distribution of
elliptical perfect electrical conductive cylinders, for various aspect ratios
$a_{x}/a_{y}$. Left panels (A, D) -- $a_{x}/a_{y}=1$; middle panels (B, E) --
$a_{x}/a_{y}=0.75$; and right panels (C, F) -- $a_{x}/a_{y}=0.5$.}}%
\end{figure}

\subsubsection*{Conditions of applicability}

The results in Figures 2 and 3 support the reliability of the effective
material parameters resulting form the \textit{supra}-classical dynamic
homogenization procedure reported here. Note however that, although not
apparent in the results, there is an approximation involved in replacing a
finite-size heterogeneous composite with its homogenized equivalent, in
addition to the reliability of the homogenization procedure (which ignores
transition region complications at the interface $\left\vert x\right\vert \leq
a$ \cite{Fikioris1964}). In practical terms, it means that for a finite sample
of the random composite the applicability of dynamic homogenization not only
depends upon the frequency under consideration but also upon the phase of the
composite at the boundary of the sample. An effort to quantify such an
approximation is described in Ref. \cite{Srivastava2014}. Here, the
approximation results from truncating interfaces of a finite (or
semi-infinite) 1-D periodic composite, the later being replaced with what are
essentially its effective dynamic properties in the infinite Bloch-wave
domain. Other questions will need to be answered relating to the shape and
size of the scattering boundary, the effect of increasing the number of
particles, and how many realizations are required to determine both the near-
and far-fields accurately. It is expected that as the bounded area increases,
so does the uncertainty of the calculated field. An investigation in this
direction is beyond the scope of the paper, however we refer the reader to the
comprehensive numerical analysis (based on the quasi-crystalline
approximation) reported in Refs. \cite{Siqueira1996, Sarabandi1997}. The later
references should come with a warning, as their analysis contains the implicit
(and incorrect \cite{Bohren1986}) assumption that the dielectric permittivity
is the only quantity of interest.

Finally, let us note that the effective material parameters derived in here
(which are tensor values for anisotropic media) are not necessarily tied to
the physical material parameters of any of the individual elements of the
metamaterial. A rather critical survey discussing the link and the difference
between these two concepts (\textit{i.e.} effective \textit{versus}
characteristic material parameters), particularly for the case of Maxwell's
equations, is presented in Ref. \cite{Simovski2010}. As evidenced in this
survey and references therein, homogenization theories continue to attract
attention and even controversy. It appears, from considering exact reflection
coefficients at oblique incidence (if one assumes that Fresnel-like formulae
are always valid), that any effective material parameters that can be
introduced in any theory would depend on the angle of incidence; broadly
speaking, they would depend on the type of illumination. This means that these
effective properties do not necessarily relate solely to the bulk properties
of the material itself; they can involve the material and the type of
illumination. Relevant considerations in this direction are presented for
periodic composites in Refs. \cite{Smigaj2008, Markel2012}. A retrieval method
extended to the arbitrary orientations of the principal axes of anisotropy and
oblique incidence was presented in Ref. \cite{Castanie2014}. A discussion
regarding modelling of the coherent wave propagation from the knowledge of the
material properties along the principal axes only is elaborated in Ref.
\cite{Caleap2013}.

To summarize, a self-consistent multiple scattering approach, which enables
the dynamic homogenization of metamaterials in two dimensions is developed.
The quasi-crystalline approximation is employed to break the hierarchy of
increasing conditional probability densities, but otherwise the treatment is
exact. In particular, the effective wavenumber and the effective impedance is
obtained. These characteristics can then be used to determine the effective
constitutive parameters of the homogenised material. Whether the resulting
effective parameters represent a true bulk property of the metamaterial in the
dynamic range is yet to be determined. The two natural approximations - dilute
media and low frequency approximations - show consistency, and, moreover, the
quasi-static limit gives results reminiscent of the laws of Maxwell Garnett
\cite{Garnett1904}, Ament \cite{Ament1953}, and Kuster and Toks\"{o}z
\cite{Kuster1974a}, respectively for electromagnetic, acoustic, and elastic
material parameters (see Ref. \cite{Supp}, section S1, for more details). The
entire analysis described in this work is germane for alternative analytical
procedures based on other scattering operators $\mathbf{Q}$\ for an isolated
particle. As shown in Ref. \cite{Supp} (section S2), a fully self-consistent
procedure may be based on a new kind of isolated scatterer problem. We have
shown that the coherent potential approximation, used in many previous works,
is only an approximation of this procedure to the first order in the
concentration of particles.

The theory provided here offers exciting opportunities for researchers in
different communities, ranging from seismic waves to the entire field of
ultrasound research, and spanning radio frequency and optical engineering. In
particular, metamaterial modelling in optics, physical acoustics, and
condensed matter physics may benefit from a rigorous, compact model for
estimating more accurate and anisotropic effective medium parameters that
homogenize artificial media.

\begin{acknowledgments}
The numerical results were obtained using the computational facilities of the
Advanced Computing Research Centre, University of Bristol - http://www.bris.ac.uk/acrc/.
\end{acknowledgments}

\renewcommand{\theequation}{A\arabic{equation}}\setcounter{equation}{0}\makeatletter\makeatother

\section*{Appendix}

\subsection{Governing equation}

There are many examples of wave equations in the physical sciences,
characterized by oscillating solutions that propagate through space and time
while, in lossless media, conserving energy. Examples include the scalar wave
equation (\textit{e.g.}, pressure waves in a gas), Maxwell's equations
(electromagnetism), Schr\"{o}dinger's equation (quantum mechanics), elastic
vibrations, and so on. From a mathematical viewpoint, all of these share
certain common features. In the following, we shall briefly identify the
similarities between three types of such waves, in two dimensions:
electromagnetic waves, anti-plane elastic waves, and acoustic waves.
Electromagnetic waves are quite different from acoustic and elastic waves in
that they can travel through vacuum. However, from an algebraic perspective,
all three types of waves can be described by a unique scalar equation and
hence these disparate phenomena can be studied simultaneously. The prototype
problem consists of the wave equation at fixed angular frequency $\omega$%

\begin{equation}
\mathbf{\nabla\cdot}\left(  \mathsf{m}^{-1}\left(  \mathbf{r}\right)
\mathbf{\nabla}\psi\left(  \mathbf{r}\right)  \right)  +\omega^{2}%
\mathsf{d}\left(  \mathbf{r}\right)  \psi\left(  \mathbf{r}\right)  =0,
\label{wave}%
\end{equation}
where $c=1/\sqrt{\mathsf{md}}$ is the phase velocity of the wave for some
parameters $\mathsf{m}\left(  \mathbf{r}\right)  $ and $\mathsf{d}\left(
\mathbf{r}\right)  $ of the medium. In the above, the scalar wavefunction
$\psi\left(  \mathbf{r}\right)  {\mathrm{e}}^{-\mathrm{i}\omega t}$
corresponds to some physical field.

The problem considered here is reduced to points in the $x$-$y$ plane
(\textit{i.e.} the cross section plane of our scattering geometry), which in
polar coordinates are $\mathbf{r=}\left(  r,\theta\right)  $; here, $\theta$
is measured from the positive $x$-axis. Let us first consider the two
important modes for electromagnetic wave propagation: the transverse electric
field and the transverse magnetic field. These modes are closely analogous to
anti-plane shear in elastodynamics and to acoustic waves. Let us assume the
medium is isotropic and has dielectric permittivity $\varepsilon$ and magnetic
permeability $\mu$ that are independent of $z.$ A transverse magnetic
(\textsc{tm}) field is a special solution of the Maxwell's equations that has
the form $\mathbf{H}=\psi\left(  x,y\right)  \mathbf{i}_{z}$, and the
electrical field $\mathrm{i}\omega\varepsilon\mathbf{E}=-\mathbf{\nabla\times
H}=\frac{\partial\psi}{\partial y}\mathbf{i}_{x}-\frac{\partial\psi}{\partial
x}\mathbf{i}_{y}.$ A transverse electric (\textsc{te}) field is another
special solution of the Maxwell's equations that has the form $\mathbf{E}%
=\psi\left(  x,y\right)  \mathbf{i}_{z}$, whereby the magnetic field is given
by $\mathrm{i}\omega\mu\mathbf{H}=\mathbf{\nabla\times E}=\frac{\partial\psi
}{\partial x}\mathbf{i}_{y}-\frac{\partial\psi}{\partial y}\mathbf{i}_{x}.$
Observe that simple knowledge of the scalar wavefunction $\psi$ suffices for
the determination of the $x$ and $y$ components of the electric and magnetic
fields, for the two polarizations. Hence, Eq. \eqref{wave}\ is the governing
wave equation for electromagnetic waves provided that $\left(  \mathsf{m}%
,\mathsf{d,}\psi\right)  \leftrightarrow\left(  \varepsilon,\mu,H_{z}\right)
$ for \textsc{tm} waves, and $\left(  \mathsf{m},\mathsf{d,}\psi\right)
\leftrightarrow\left(  \mu,\varepsilon,E_{z}\right)  $ for \textsc{te} waves.
The general solution independent of $z$ is a superposition of the \textsc{te}
and \textsc{tm} solutions. This can be seen by observing that the Maxwell's
equations decouple under this condition and a general solution can be written
as $\left(  H_{x},H_{y},H_{z}\right)  =\left(  H_{x},H_{y},0\right)  +\left(
0,0,H_{z}\right)  $, where the second term represents the \textsc{tm}
solution. The first term is of course the \textsc{te} solution because
$\mathbf{\nabla\times}\left(  H_{x},H_{y},0\right)  =(0,0,\tfrac{\partial
}{\partial y}H_{x}-\frac{\partial}{\partial x}H_{y})$, which implies
$E_{x}=E_{y}=0$ as expected. Let us now consider the case of anti-plane shear
strain which is a special state of deformation where the displacement field is
given by $\mathbf{u}=\psi\left(  x,y\right)  \mathbf{i}_{z}.$ This is an
out-of-plane mode of deformation and is analogous to transverse
electromagnetic wave propagation. In the linear regime, an isotropic elastic
medium is characterized by its density $\rho$ and the Lam\'{e} elastic
constants $G\equiv\mu$ and $\lambda$: $G$ is the shear modulus (notation used
to distinguish from the permeability $\mu$ employed in electromagnetics) and
$\lambda=$ $\kappa-G$ where $\kappa$ is the two dimensional bulk modulus.
Hence, Eq. \eqref{wave}\ is the wave equation for anti-plane shear provided
that $\left(  \mathsf{m},\mathsf{d,}\psi\right)  \leftrightarrow\left(
1/G,\rho,u_{z}\right)  $. Shear waves that satisfy this equation are also
called \textsc{sh} (shear horizontal) waves, particularly in seismology.
Finally, let us consider the acoustic wave propagation in an isotropic medium.
For an inviscid fluid or gas, the shear modulus $G$ is zero -- and $\lambda$
is just the bulk modulus. In this case, replacing $\psi$ with the pressure
$p=-\kappa\mathbf{\nabla}\cdot\mathbf{u}$, we obtain precisely the acoustic
wave equation \eqref{wave}\ for \textsc{p} (or compressional) waves if
$\left(  \mathsf{m},\mathsf{d}\right)  \leftrightarrow\left(  \rho
,1/\kappa\right)  $.

In essence, the solutions of the three problems considered will lead to
similar conclusions if we make the appropriate interpretation of the
quantities involved.

\renewcommand{\theequation}{B\arabic{equation}}\setcounter{equation}{0}\makeatletter\makeatother

\subsection{Effective field method}

Suppose that discrete particles of cylindrical geometry are randomly and
uniformly distributed in a half-space defined by $\left\{  x>0\right\}  $. The
particles need not be circular, provided that each of them can be contained in
a circumscribing circular surface of radius $a_{j}$ (with an axis of
revolution parallel to the $z$-axis); their number density is $\eta_{j}$. Both
the particles and the matrix are made of isotropic materials. Let a plane wave
${\psi}=\mathrm{\exp}\left[  \mathrm{i}\left(  kx-\omega t\right)  \right]  $
of unit amplitude propagate with wavenumber $k$ in the matrix along the
$x$-direction. When this wave propagates in the composite material, multiple
scattering occurs. Either propagation or diffusion, or a combination of the
two phenomena is observed, depending on the frequency as well as on the
geometrical and material properties of the composite. Assuming that
propagation occurs, one can describe the coherent wave motion in the composite
by a complex-valued wavenumber $\mathcal{K}$. The fundamental equation for
configurational averages of the exciting and total fields for scalar
wavefunctions has been derived in detail in Refs.
\cite{Foldy1945,Fikioris1964,Siqueira1996,Linton2005}. The quasi-crystalline
approximation \cite{Lax1952} is used to truncate the hierarchy of equations
(Foldy-Lax hierarchy) so that only the correlation between every two particles
is considered. We obtain the implicit dispersion equation for the effective
wavenumber $\mathcal{K}$\ of the coherent wave $\mathrm{\exp}\left[
\mathrm{i}\left(  \mathcal{K}x-\omega t\right)  \right]  $,%

\begin{equation}
\mathcal{K}^{2}=k^{2}+%
{\textstyle\sum\nolimits_{j}}
\epsilon_{j}{{\mathcal{F}}}\left(  a_{j}\right)  , \label{keff}%
\end{equation}
where $\epsilon_{j}=4\pi\eta_{j}$, and the effective scattering amplitude
${{{\mathcal{F}}}}$ is given by%

\begin{equation}
{{\mathcal{F}}}\left(  a_{j}\right)  ={{\mathbf{e}}}^{t}\big ({{{\mathbf{Q}}%
}_{j}^{-1}-%
{\textstyle\sum\nolimits_{i}}
\epsilon_{i}{{{\mathbf{R}}}_{ij}}}\big )^{-1}{\mathbf{e}}, \label{4)}%
\end{equation}
${\mathbf{e}}=\left(  1,1,\dots\right)  ^{t}$ is a constant unit vector. The
shorthand notation ${{\mathbf{R}}}_{ij}\equiv{\mathbf{R}}\left(
b_{ij}\right)  $ and ${{\mathbf{Q}}}_{j}\equiv{\mathbf{Q}}\left(
a_{j}\right)  $ has been used. The infinite square matrices ${\mathbf{R}%
}\left(  b_{ij}\right)  $ and ${\mathbf{Q}}\left(  a_{j}\right)  $, have elements%

\begin{equation}
{{\mathcal{R}}}_{n\nu}\left(  b_{ij}\right)  =\frac{\mathcal{P}_{n-\nu}\left(
\mathcal{K}b_{ij}\right)  -1}{\mathcal{K}^{2}-k^{2}}{\mathbf{+}}{{\mathcal{N}%
}}_{n-\nu}\left(  \mathcal{K}b_{ij}\right)  , \label{R}%
\end{equation}
where $b_{ij}\geqq a_{i}+a_{j}$, and%

\begin{equation}
Q_{n\nu}\left(  a_{j}\right)  =\frac{\mathrm{1}}{\mathrm{i}\pi}{\delta_{n\nu
}T}_{n}\left(  a_{j}\right)  . \label{6)}%
\end{equation}
Here, $\delta_{ij}$ denotes the Kronecker delta,\ and $\mathcal{P}_{\ell}$ and
${{\mathcal{N}}}_{\ell}$ are given by%

\begin{align*}
\mathcal{P}_{\ell}\left(  z\right)   &  =\frac{\mathrm{i}\pi}{2}\left[
zH_{\ell}^{(1)}\left(  x\right)  \tfrac{\mathrm{d}}{\mathrm{d}z}J_{\ell
}\left(  z\right)  -xJ_{\ell}\left(  z\right)  \tfrac{\mathrm{d}}{\mathrm{d}%
x}H_{\ell}^{(1)}\left(  x\right)  \right]  ,\\
{{\mathcal{N}}}_{\ell}\left(  z\right)   &  =\frac{\mathrm{i}\pi}{2}%
\int_{b_{ij}}^{\infty}{\mathrm{d}r\left[  g_{ij}\left(  r\right)  -1\right]
}{rH}_{\ell}^{(1)}\left(  kr\right)  J_{\ell}\left(  {zr}/{b_{ij}}\right)  ,
\end{align*}
where $J_{\ell}$ and $H_{\ell}^{(1)}$ are the cylindrical Bessel and Hankel
functions, respectively, and $x=kb_{ij}.$ The function $g_{ij}$ is the
cross-pair distribution function of two particle species (with sizes $a_{i}$
and $a_{j}$), and satisfies the non-overlapping condition: $g_{ij}\left(
r\right)  =0$ for $r<b_{ij}$; also, if the distance between particles tends to
infinity, then the correlation between their locations disappears,
\textit{i.e.}, $\underset{r\rightarrow\infty}{\lim}g_{ij}\left(  r\right)
=1$. The scattering coefficients $T_{n}\left(  a_{j}\right)  $ in Eq.
\eqref{6)} depend on frequency, size $a_{j}$, as well as on the properties of
the particle and those of the matrix material; they are evaluated by imposing
appropriate boundary conditions at $r=a_{j}$.

Equation \eqref{keff} follows directly from a Lorentz--Lorenz-type law, and is
an exact expression for the effective wavenumber, subject to the
quasi-crystalline approximation. It is of interest to note how various
physical aspects are embedded in this equation. The scattering matrix
${{{\mathbf{Q}}}}$ describes the response of a single particle to a plane
incident harmonic wave with wavenumber $k$, and contains all the scattering
behaviour in terms of particle geometry and physical parameters. The effective
wavenumber $\mathcal{K}$ only appears in the matrix ${{{\mathbf{R}}}}$, which
is defined by the spatial arrangements of particles, and accounts for multiple
scattering. Should the distribution of particles be regular, the
quasi-crystalline approximation is exact, in which case the
multiple-scattering matrix ${{{\mathbf{R}}}}$\ can be reduced to a well known
lattice sum.

The theory described above is now complete insofar as behaviour within the
medium\ is concerned. It is also of interest, however, to calculate the
effective impedance $\mathcal{Z}$, which defines the reflectivity of the
half-space $\left\{  x>0\right\}  $ - a quantity which may be measured
directly. Following the derivations in Refs. \cite{Fikioris1964, Caleap2012},
the coherent reflected field $\left\langle {\psi}\right\rangle =\mathfrak{R}%
\exp(-\mathrm{i}kx)$ at the half-space boundary can be obtained explicitly,
with the reflection coefficient defined as%

\begin{equation}
\mathfrak{R=}\frac{\mathfrak{-}\sum_{j}\epsilon_{j}{{\mathcal{F}}}_{\pi
}\left(  a_{j}\right)  }{4k^{2}+\sum_{j}\epsilon_{j}{{\mathcal{F}}}_{0}\left(
a_{j}\right)  }. \label{refl}%
\end{equation}
Here, $\mathfrak{R}$ represents the average (coherent) back-scattered
amplitude at normal incidence in the domain $\{x<0\}$. The effective
scattering amplitudes, ${{\mathcal{F}}}_{0}$ and ${{\mathcal{F}}}_{\pi}$,
correspond to coherent waves scattered in the forward and backward directions,
respectively, and are given by%
\begin{equation}
{{\mathcal{F}}}_{0}\left(  a_{j}\right)  ={{\mathbf{e}}}^{t}{{{\mathbf{Q}}%
}_{j}}\mathbf{v}_{j}\text{ and }{{\mathcal{F}}}_{\pi}\left(  a_{j}\right)
={{\mathbf{e}}}^{t}{{\mathbf{J}{\mathbf{Q}}}_{j}}\mathbf{v}_{j},
\end{equation}
where ${{\mathbf{J}}}=\left\{  {\delta_{n\nu}\cos n\pi}\right\}  $ is a
diagonal infinite matrix. The infinite eigenvector $\mathbf{v}_{j}$,
associated with the wavenumber equation, follows from an Ewald--Oseen-type
extinction theorem, with the result%

\begin{equation}
\mathbf{v}_{j}=\frac{2k}{\mathcal{K}+k}\big ({{\mathbf{I}}-%
{\textstyle\sum\nolimits_{i}}
\epsilon_{i}{{\mathbf{Q}}}_{j}{{{\mathbf{R}}}_{ij}}}\big )^{-1}{\mathbf{e,}}%
\end{equation}
where ${{\mathbf{I}}}$ is a unit infinite matrix. Martin \cite{Martin2011} has
obtained a formula for $\mathfrak{R}$ for obliquely incident waves on a
half-space of circular scatterers; it can be shown that at normal incidence
the result in his Eq. (39) gives agreement with Eq. \eqref{refl}. The
behaviour of the fields across interfaces was also examined in Refs.
\cite{Aguiar2006,Aristegui2010, Martin2011}; It was found that the fields
themselves are continuous but the slopes are discontinuous. Using the estimate
for the slope discontinuity, effective constitutive parameters can be derived,
as shown in Refs. \cite{Aristegui2010, Martin2011}. Equation \eqref{refl} is
an exact formula for the reflection coefficient\footnote{Note however that
during the derivation, complications in the transition region $-a\leq x\leq a$
near and on both sides of the interface have been ignored \cite{Fikioris1964}%
.}; it can be used to determine effective parameters ($\mathsf{m}%
_{eff},\mathsf{d}_{eff}$) uniquely. It is often assumed that the effective
medium corresponding to the distribution of particles may be described as a
homogeneous medium from the standpoint of coherent wave propagation -- the
homogenized equivalent having the effective dynamic properties of the
composite. In the following, we shall use this analogy, whereby the reflection
coefficient $\mathfrak{R}$ at the interface between the homogeneous medium and
the homogenized equivalent, may be written (as is standard) in terms of
impedances (resulting in a Fresnel-like formula).\footnote{For instance, for
acoustic waves $\mathfrak{R=}\left(  {{\mathcal{Z}}}-z\right)  /\left(
{{\mathcal{Z}}}+z\right)  $: this result implies the continuity of pressure
and normal velocity at the interface; for anti-plane elastic waves
$\mathfrak{R=-}\left(  {{\mathcal{Z}}}-z\right)  /\left(  {{\mathcal{Z}}%
}+z\right)  $: here, the continuity of the out-of-plane displacement and the
corresponding stress are implicit. Similar results in electromagnetics are
known as Fresnel relations (for \textsc{te} and \textsc{tm} waves).} Then,
equating the result with Eq. \eqref{refl}, the effective impedance
$\mathcal{Z}$ can be explicitly calculated. As expected, the effective
impedance is different for different polarizations. Two cases are possible,
with the following results:%

\begin{equation}
{{\mathcal{Z}}}^{\mathsf{m}}={{\mathcal{Z}}}/z\text{ and }{{\mathcal{Z}}%
}^{\mathsf{d}}=z/{{\mathcal{Z}}}, \label{zmd}%
\end{equation}
where%

\begin{equation}
{{\mathcal{Z}}}=z\frac{4k^{2}+\sum_{j}\epsilon_{j}\left[  {{\mathcal{F}}}%
_{0}\left(  a_{j}\right)  -{{\mathcal{F}}}_{\pi}\left(  a_{j}\right)  \right]
}{4k^{2}+\sum_{j}\epsilon_{j}\left[  {{\mathcal{F}}}_{0}\left(  a_{j}\right)
+{{\mathcal{F}}}_{\pi}\left(  a_{j}\right)  \right]  }. \label{zz}%
\end{equation}
Here, $z=\sqrt{\mathsf{m}/\mathsf{d}}$ is the impedance of the matrix; the
superscripts \text{`}$\mathsf{m}$'\ and `$\mathsf{d}$'\ correspond to
different physical situations, as we shall see below. We can now state our
most general expressions for the effective dynamic constitutive parameters
($\mathsf{m}_{eff},\mathsf{d}_{eff}$),
\begin{equation}
\frac{\mathsf{m}_{eff}}{\mathsf{m}}=\frac{\mathcal{K}}{k}{{\mathcal{Z}}%
}^{\mathsf{m}}\text{ and }\frac{\mathsf{d}_{eff}}{\mathsf{d}}=\frac
{\mathcal{K}}{k}{{\mathcal{Z}}}^{\mathsf{d}} \label{mdeff}%
\end{equation}
where (${{\mathcal{Z}}}^{\mathsf{m}},{{\mathcal{Z}}}^{\mathsf{d}}$) are
defined in Eqs. \eqref{zmd}-\eqref{zz}. Observe that, by using the
definition,
\begin{equation}
\mathcal{K}=k+\frac{1}{2k}%
{\textstyle\sum\nolimits_{j}}
\epsilon_{j}{{{\mathcal{F}}}}_{0}\left(  a_{j}\right)
\end{equation}
in Eq. \eqref{mdeff}, the resulting parameters ($\mathsf{m}_{eff}%
,\mathsf{d}_{eff}$) can be expressed explicitly in terms of the effective
forward and back-scattering shape functions, ${{\mathcal{F}}}_{0}$ and
${{\mathcal{F}}}_{\pi}$.

To conclude this section, we consider the line-like approximation of the
constitutive parameters \eqref{mdeff}. For this, the size of the particles is
assumed small compared to the incident wavelength ($a_{j}\ll\Lambda$). At
leading order, the single-scattering operator $\mathbf{Q}$ is compact and has
only three eigenvalues of finite size (related to terms with $n=0,\pm1$).
Furthermore, the infinite multiple-scattering operator $\mathbf{R}$ is reduced
to a rank 3 matrix. Omitting the details, we find for circular cylinders (with
$T_{1}=T_{-1}$),%

\begin{align}
\frac{\mathsf{m}_{eff}}{\mathsf{m}}  &  \simeq\frac{k^{2}+\frac{1}%
{\mathrm{i}\pi}\sum_{j}\epsilon_{j}T_{1}\left(  a_{j}\right)  }{k^{2}-\frac
{1}{\mathrm{i}\pi}\sum_{j}\epsilon_{j}T_{1}\left(  a_{j}\right)  }%
\text{,}\label{me}\\
\frac{\mathsf{d}_{eff}}{\mathsf{d}}  &  \simeq1+\frac{1}{\mathrm{i}\pi k^{2}}%
{\textstyle\sum\nolimits_{j}}
\epsilon_{j}T_{0}\left(  a_{j}\right)  \text{.} \label{de}%
\end{align}

It can be shown that\ in the quasi-static limit ($a_{j}\ll\Lambda_{0}$) the
effective property $\mathsf{m}_{eff}$ is reminiscent of the laws of Maxwell
Garnett \cite{Born1999}, Ament \cite{Ament1953}, and Kuster and Toks\"{o}z
\cite{Kuster1974a}, in two dimensions, respectively for electromagnetic,
acoustic, and elastic material parameters. (This is further described in Ref.
\cite{Supp}, section S1.) On the other hand, the effective property
$\mathsf{d}_{eff}$ reduces to the simple and inverse rules of mixtures,
depending on the physical model under consideration, and as seen from Eq.
\eqref{de} is linear in $\epsilon_{j}$.

\renewcommand{\theequation}{C\arabic{equation}}\setcounter{equation}{0}\makeatletter\makeatother

\subsection{Explicit second order approximations}

At low concentrations ($\epsilon_{j}a_{j}^{2}\ll1$), the dispersion equation
is explicit, and reduces to the well-known formula \cite{Foldy1945}%

\begin{equation}
\mathcal{K}^{2}\simeq k^{2}+%
{\textstyle\sum\nolimits_{j}}
\epsilon_{j}f_{0}\left(  a_{j}\right)  ,
\end{equation}
where the forward-scattering amplitude $f_{0}$ is given by $f_{0}\left(
a_{j}\right)  ={{\mathbf{e}}}^{t}{{\mathbf{Q}}}_{j}{\mathbf{e}}.$ More
generally, the angular shape function $f_{\theta}$ for each particle is
defined, in terms of Fourier series, as\footnote{Note the shorthand notation $%
{\textstyle\sum\limits_{n}}
=%
{\textstyle\sum\limits_{n=-\infty}^{\infty}}
$ is used throughout.}%

\begin{equation}
f_{\theta}\left(  a_{j}\right)  =\frac{\mathrm{1}}{\mathrm{i}\pi}\sum
_{n}{T_{n}\left(  a_{j}\right)  {\mathrm{e}}^{{{\mathrm{i}n\theta}}}.}
\label{ftheta}%
\end{equation}

An expansion of the dispersion equation \eqref{keff} to the second order in
concentration results in%

\begin{equation}
\mathcal{K}^{2}\simeq k^{2}+%
{\textstyle\sum\nolimits_{j}}
\epsilon_{j}{{\mathbf{e}}}^{t}\mathbf{Q}_{j}\mathbf{e}+{%
{\textstyle\sum\nolimits_{i,j}}
}\epsilon_{i}\epsilon_{j}{{\mathbf{e}}}^{t}{\mathbf{Q}}_{i}%
\widetilde{{\mathbf{R}}}_{ij}{\mathbf{Q}}_{i}\mathbf{e}+{\mathcal{O}}\left(
\epsilon_{i}\epsilon_{j}\epsilon_{k}\right)  , \label{keff0}%
\end{equation}
where the matrix $\widetilde{{\mathbf{R}}}_{ij}\equiv\widetilde{{\mathbf{R}}%
}\left(  {b_{ij}}\right)  =\underset{\mathcal{K}\rightarrow k}{\lim
}{{\mathbf{R}}}_{ij}$ and has elements%

\begin{align}
\widetilde{{{\mathcal{R}}}}_{n\nu}\left(  {b_{ij}}\right)   &  ={{\mathcal{N}%
}}_{\ell}\left(  x\right)  +\frac{\mathrm{i}\pi}{4k^{2}}\big [\left(  \ell
^{2}-x^{2}\right)  J_{\ell}\left(  x\right)  H_{\ell}^{(1)}\left(  x\right)
\nonumber\\
&  -x^{2}\tfrac{\mathrm{d}}{\mathrm{d}x}J_{\ell}\left(  x\right)
\tfrac{\mathrm{d}}{\mathrm{d}x}H_{\ell}^{(1)}\left(  x\right)  \big ],
\label{reff}%
\end{align}
with $\ell=n-\nu$, and $x=kb_{ij}$. Note that for spatially uncorrelated
particles, ${{\mathcal{N}}}_{\ell}\left(  x\right)  =0.$

For the effective impedance of Eq. \eqref{zz}, at first order in
concentration, we have%

\begin{equation}
\mathcal{Z}=z-\frac{1}{2k^{2}}%
{\textstyle\sum\nolimits_{j}}
\epsilon_{j}f_{\pi}\left(  a_{j}\right)  ,
\end{equation}
where the back-scattering amplitude $f_{\pi}$ is given by $f_{\pi}\left(
a_{j}\right)  ={{\mathbf{e}}}^{t}\mathbf{J}{{\mathbf{Q}}}_{j}{\mathbf{e}}$.
The second order approximation is too long to warrant including here. For
completeness, we also give the following results, in terms of Fourier series,
\begin{equation}
{{\mathbf{e}}}^{t}{\mathbf{Q}}_{i}\widetilde{{\mathbf{R}}}_{ij}\mathbf{Q}%
_{j}\mathbf{e=}\frac{\mathrm{i}}{\pi^{2}}\sum_{n,\upsilon}%
\widetilde{{{{\mathcal{R}}}}}{_{n\nu}}\left(  b_{ij}\right)  {T_{n}\left(
a_{i}\right)  T_{\nu}\left(  a_{j}\right)  },
\end{equation}%
\begin{equation}
{{\mathbf{e}}}^{t}{\mathbf{JQ}}_{i}\widetilde{{\mathbf{R}}}_{ij}\mathbf{Q}%
_{j}\mathbf{e=}\frac{\mathrm{i}}{\pi^{2}}\sum_{n,\upsilon}{{\left(  -1\right)
}^{n}}\widetilde{{{{\mathcal{R}}}}}{_{n\nu}}\left(  b_{ij}\right)
{T_{n}\left(  a_{i}\right)  T_{\nu}\left(  a_{j}\right)  .}%
\end{equation}

These expressions can be easily approximated in the low frequency limit by
observing that, to leading order in ($kb_{ij}$), and for uncorrelated
particles, $\widetilde{{{\mathcal{R}}}}_{n\nu}\cong\left\vert n-\nu\right\vert
/2k^{2}$. The results obtained here have been used to derive the analytic
formulae presented in the main text.

\widetext
\clearpage
\balancecolsandclearpage

\textbf{\large Supplementary text}

\setcounter{equation}{0}
\setcounter{figure}{0}
\setcounter{table}{0}
\setcounter{page}{1}
\makeatletter
\renewcommand{\theequation}{S\arabic{equation}}
\renewcommand{\thefigure}{S\arabic{figure}}
\renewcommand{\bibnumfmt}[1]{[S#1]}
\renewcommand{\citenumfont}[1]{S#1}
\makeatother

\subsection*{S1. Consistency check}

It is straightforward to expand the low-frequency formulae obtained in the
Appendix. Instead, the expansions (1) and (2) [or (7) and (8)] are
approximated for long wavelengths ($a\ll\Lambda$).

Small cylinders behave as a combination of a monopole (or a source) and a
dipole: this is a generic situation, and the leading order contribution to the
angular shape function $f_{\theta}$ only involves the scattering coefficients
$T_{0}\left(  \omega\right)  $ and $T_{1}\left(  \omega\right)  $,
\textit{i.e.}
\begin{equation}
f_{\theta}\simeq\frac{1}{\mathrm{i}\pi}\left(  T_{0}+2T_{1}\right)  \cos
\theta.
\end{equation}

The integrals in Eq. (6) reduce to
\begin{equation}%
\mathscr{H}%
_{0}=\frac{8}{\pi^{2}}T_{1}\left(  T_{0}+T_{1}\right)  \text{ and }%
\mathscr{H}%
_{\pi}=-\frac{8}{\pi^{2}}T_{1}^{2}.
\end{equation}

Finally, using these results, the effective parameters may then be expressed as%

\begin{align}
\frac{\mathsf{m}_{eff}}{\mathsf{m}}  &  \simeq1+\epsilon\frac{2}{\mathrm{i}\pi
k^{2}}T_{1}-\epsilon^{2}\frac{2}{\pi^{2}k^{4}}T_{1}^{2},\label{m00}\\
\frac{\mathsf{d}_{eff}}{\mathsf{d}}  &  \simeq1+\epsilon\frac{1}{\mathrm{i}\pi
k^{2}}T_{0}, \label{d00}%
\end{align}
which agree to\ $\mathcal{O}\left(  \epsilon^{2}\right)  $ with the general
estimates (B12) and (B13) in Appendix B, as expected. We can see that
$\mathsf{d}_{eff}$ and $\mathsf{m}_{eff}$ are related to monopolar ($n=0$) and
dipolar ($n=1$) scattering coefficients, respectively. Note that the results
in Eqs. \eqref{m00} and \eqref{d00} need not correspond to the quasi-static
limit, because wavelength (and $1/\mathrm{e}$ length, if losses are present)
within the particles is as yet arbitrary relative to the particle size. Many
previous effective medium results
\cite{Li2004,Wu2006,Wu2007,Jin2009a,Jin2012,Zhang2015} correspond to such
dynamic approximations. If we further assume that the wavelength within the
particles is small ($a\ll\Lambda_{0}$), it can be shown that to the leading
order in $(ka)$,%

\begin{equation}
T_{0}\simeq\frac{\mathrm{i}\pi}{4}k^{2}a^{2}\mathcal{D}\text{ and }T_{1}%
\simeq\frac{\mathrm{i}\pi}{4}k^{2}a^{2}\mathcal{M}, \label{tt}%
\end{equation}
where%

\begin{equation}
\mathcal{D}=\frac{\mathsf{d}_{0}}{\mathsf{d}}-1\text{ and }\mathcal{M}%
=\frac{\mathsf{m}_{0}-\mathsf{m}}{\mathsf{m}_{0}+\mathsf{m}}. \label{dm}%
\end{equation}

The quasi-static relations \eqref{tt} with coefficients \eqref{dm} hold for a
variety of boundary value problems, equivalent across electromagnetics,
acoustics, and elasticity. This includes the Neumann boundary condition by
setting $\mathsf{d}_{0}=\mathsf{m}_{0}=0$. As is well known, the Dirichlet
condition is atypical, and special care is needed.\footnote{A particular
feature of the Dirichlet problem is the presence of $\log ka$ in the
asymptotics of the solution. The interpretation of this effect depends on the
physical model under consideration. For investigations in this direction, see,
\textit{e.g.}, Ref. \cite{Movchan2001}.} In electromagnetism, Dirichlet or
Neumann boundary conditions (depending on the polarization in question)
describe inclusions that are perfectly conducting.

In terms of $\mathcal{D}$ and $\mathcal{M}$, Eqs. \eqref{m00} and \eqref{d00} yield%

\begin{align}
\frac{\mathsf{m}_{eff}}{\mathsf{m}}  &  \simeq1+2\phi\mathcal{M+}2\phi
^{2}\mathcal{M}^{2},\label{mdd}\\
\frac{\mathsf{d}_{eff}}{\mathsf{d}}  &  \simeq1+\phi\mathcal{D},
\end{align}
where $\phi=\epsilon a^{2}/4$ ($=\pi\eta a^{2})$ denotes the fractional volume
occupied by the particles. These equations reduce to different forms in the
quasi-static limit depending on application. Let us now specialize them to
electromagnetic, acoustic, and elastic scattering, in succession.

\subsubsection*{Electromagnetic waves}

There is a vast literature describing the many approaches to calculate
effective-medium electromagnetic parameters \cite{Waterman1986}, and many of
the existing theories are closely related to models developed in the late
1800s and early 1900s. We note in particular the expressions for the effective
permittivity $\varepsilon_{eff}$\ obtained in three dimensions by Maxwell
Garnett \cite{Garnett1904} and by Bruggeman \cite{Bruggeman1935} that, in
turn, are closely related to the older Lorentz--Lorenz formula for
time-dependent electric fields and the Clausius--Mosotti equation for static
fields \cite{Born1999}. The Bruggeman formula has the special property that it
treats the particles and the environment symmetrically. However, this results
in a quadratic order in $\phi$\ that is different from the expansion of
Maxwell Garnett rule. Conversely, our result for $\varepsilon_{eff}$ is
consistent with the rule of Maxwell Garnett. Indeed, consider a mixture where
small magnetoelectric particles are embedded in a host environment of
permittivity (permeability) $\varepsilon$ ($\mu$). The complex permittivity of
the particles is $\varepsilon_{0}$ and their permeability is $\mu_{0}$. Then,
for \textsc{tm} waves, we have%

\begin{align}
\varepsilon_{eff}^{\text{\textsc{tm}}}  &  \simeq\varepsilon+2\phi
\varepsilon\frac{\varepsilon_{0}-\varepsilon}{\varepsilon_{0}+\varepsilon
}\mathcal{+}2\phi^{2}\varepsilon\left(  \frac{\varepsilon_{0}-\varepsilon
}{\varepsilon_{0}+\varepsilon}\right)  ^{2},\label{eps}\\
\mu_{eff}^{\text{\textsc{tm}}}  &  \simeq\mu+\phi\left(  \mu_{0}-\mu\right)  .
\label{mu}%
\end{align}

Should the incident wave be electric in nature, $\varepsilon$ and $\mu$ would
have been interchanged. In the language of the dielectric problem, Eq.
\eqref{eps} is reminiscent of the Maxwell Garnett estimate (more precisely, a
small-$\phi$ approximation of the Maxwell Garnett rule in two dimensions). On
the other hand, the effective permeability of Eq. \eqref{mu} is given by a
simple rule of mixtures.

\subsubsection*{Sound waves in a compressible fluid}

In the acoustics context, consider a fluid-particle mixture and let
$(\rho,\kappa)$ and $(\rho_{0},\kappa_{0})$ be mass densities and bulk moduli
of their respective phases. The particle material can be, \textit{e.g.} solid
or fluid. Then,%

\begin{align}
\rho_{eff}^{\text{\textsc{p}}}  &  \simeq\rho+2\phi\rho\frac{\rho_{0}-\rho
}{\rho_{0}+\rho}\mathcal{+}2\phi^{2}\rho\left(  \frac{\rho_{0}-\rho}{\rho
_{0}+\rho}\right)  ^{2},\label{rho}\\
\kappa_{eff}  &  \simeq\left(  \frac{1-\phi}{\kappa}+\frac{\phi}{\kappa_{0}%
}\right)  ^{-1}. \label{kappa}%
\end{align}

The mass density $\rho_{eff}^{\text{\textsc{p}}}$ is analogue to that obtained
in three dimensions by Ament \cite{Ament1953} (more precisely, a small-$\phi$
approximation of the Ament-estimate in two dimensions). Equation \eqref{kappa}
is recognized as the Reuss average for the effective bulk modulus
$\kappa_{eff}$.

\subsubsection*{Anti-plane elastic waves in a solid composite}

Consider the elastodynamic problem of anti-plane shear scattering, and let the
host and particles have shear moduli $G$ and $G_{0}$, respectively. Again, the
particles can be made of, \textit{e.g.} solid or fluid material. For solid
particles, we obtain%

\begin{align}
\frac{G}{G_{eff}}  &  \simeq1+2\phi\frac{G-G_{0}}{G+G_{0}}\mathcal{+}2\phi
^{2}\left(  \frac{G-G_{0}}{G+G_{0}}\right)  ^{2},\\
\rho_{eff}^{\text{\textsc{sh}}}  &  \simeq\rho+\phi\left(  \rho_{0}%
-\rho\right)  .
\end{align}

The effective shear modulus $G_{eff}$ is analogue to that obtained in three
dimensions by Kuster and Toks\"{o}z \cite{Kuster1974a} (more precisely, a
small-$\phi$ approximation of the Kuster-Toks\"{o}z estimate in two
dimensions). Observe that the effective mass density $\rho_{eff}%
^{\text{\textsc{sh}}}$ in the case of anti-plane elasticity is also given by a
simple rule of mixtures.

\renewcommand{\thesection}{S\arabic{section}}\makeatother

\subsection*{S2. Self-consistent effective field method}

Self-consistent methods for the problem of scalar wave propagation through a
medium with many particles may be found in the works of Maxwell and Rayleigh.
During more than a century, in a number of works, these methods were
extensively developed and used for the solution of various wave propagation
problems. The present general results may be recast in terms of a
\textit{dual-layer} scattering operator ${{\mathbf{Q}}}_{eff}$ corresponding
to a coated particle embedded in an effective medium with the properties
$\mathsf{m}_{eff}$ and $\mathsf{d}_{eff}$. The coating is made of the original
matrix material with the properties $\mathsf{m}$ and $\mathsf{d}$. For
simplicity, we assume the particles are circular cylinders and have equal
sizes $a_{j}=a$. The radius $c$ of their coating defines the volume fraction
to be $\phi=a^{2}/c^{2}$. Note that the theory presented below is not limited
to only these geometries but applicable, in principle, to any other particle
shapes. From Eq. (B1), we consider the effective wavenumber at the second
order in concentration as,
\begin{equation}
\mathcal{K}^{2}=k_{eff}^{2}+\epsilon{{\mathbf{e}}}^{t}{{\mathbf{Q}}}%
_{eff}{\mathbf{e}}+\epsilon^{2}{{\mathbf{e}}}^{t}{{\mathbf{Q}}}_{eff}%
\widetilde{{{\mathbf{R}}}}_{eff}{{\mathbf{Q}}}_{eff}{\mathbf{e}+\mathcal{O}%
}\left(  \epsilon^{3}\right)  , \label{self}%
\end{equation}
where the effective operator ${{\mathbf{Q}}}_{eff}$ corresponds to a coated
particle excited by the coherent motion $\exp(\mathrm{i}k_{eff}x)$. The
multiple-scattering matrix $\widetilde{{{\mathbf{R}}}}_{eff}$ is given by Eq.
(C4), but with $k$ replaced by $k_{eff}$. The self-consistent scheme now
assumes that $\mathcal{K}=k_{eff}$. From a physical point of view this means
that the coherent wavefield the composite medium coincides with the wavefield
propagating in the effective medium. Hence, the medium can be considered as
homogenized since there is no scattering in the outer effective medium. This
results in the following non-linear equation for $k_{eff}$,%

\begin{equation}
{{\mathbf{e}}}^{t}{{\mathbf{Q}}}_{eff}{\mathbf{e}}={\epsilon{\mathbf{e}}}%
^{t}{{\mathbf{Q}}}_{eff}\widetilde{{{\mathbf{R}}}}_{eff}{{\mathbf{Q}}}%
_{eff}{\mathbf{e}+\mathcal{O}}\left(  \epsilon^{2}\right)  . \label{sqca}%
\end{equation}
It is thought that the use of the self-consistent scheme \eqref{sqca} applied
to the effective wavenumber (B1) can improve the accuracy of the results while
the concentration of particles increases. Note that since ${{\mathbf{Q}}%
}_{eff}$\ is a transcendental function of the unknown $k_{eff}$, explicit
solutions can only be obtained subject to low-frequency approximation. Writing
the dual-layer forward-scattering amplitude $f_{0}^{eff}$ of a coated
particle, as $f_{0}^{eff}={{\mathbf{e}}}^{t}{{\mathbf{Q}}}_{eff}{\mathbf{e}}$,
Eq. \eqref{sqca} is then reduced to%

\begin{equation}
f_{0}^{eff}\cong-\frac{2{\epsilon}}{\pi^{2}k_{eff}^{2}}T_{1}^{eff}\left(
T_{0}^{eff}+T_{1}^{eff}\right)  . \label{f0eff}%
\end{equation}

The effective scattering coefficients $T_{0}^{eff}$ and $T_{1}^{eff}$ can be
calculated in terms of $\phi$ ($=\epsilon a^{2}/4$), as%

\begin{align}
T{_{0}^{eff}}  &  {\simeq}\frac{\mathrm{i}\pi}{4\phi}k_{eff}^{2}%
a^{2}\bigg [\frac{\mathsf{d}}{\mathsf{d}_{eff}}\left(  1+\phi\mathcal{D}%
\right)  -1\bigg ],\label{T0}\\
{T_{1}^{eff}}  &  {\simeq}\frac{\mathrm{i}\pi}{4\phi}k_{eff}^{2}a^{2}%
\frac{\left(  1+\phi\mathcal{M}\right)  \mathsf{m}-\left(  1-\phi
\mathcal{M}\right)  \mathsf{m}_{eff}}{\left(  1+\phi\mathcal{M}\right)
\mathsf{m+}\left(  1-\phi\mathcal{M}\right)  \mathsf{m}_{eff}}, \label{T11}%
\end{align}
\newline where $\mathcal{M}$ and $\mathcal{D}$ are defined in Eq. \eqref{dm}.
We find directly that $T_{1}^{eff}=\mathcal{O}\left(  \phi^{3}\right)  $, if
one uses the small-$\phi$ estimate \eqref{mdd}. More generally, without
restriction on $\phi$, we obtain $T_{0}^{eff}=T_{1}^{eff}=0$, for the
following effective parameters,%

\begin{align}
\frac{\mathsf{m}_{eff}}{\mathsf{m}}  &  =\frac{1+\phi\mathcal{M}}%
{1-\phi\mathcal{M}},\label{meffx}\\
\frac{\mathsf{d}_{eff}}{\mathsf{d}}  &  =1+\phi\mathcal{D}. \label{deffx}%
\end{align}
Here, $\mathsf{m}_{eff}$ and $\mathsf{d}_{eff}$ are the quasi-static limits of
Eqs. (B12) and (B13). This result is significant, since \eqref{f0eff} reduces
to the \textit{coherent potential approximation }of solid state physics,
regularly used in electromagnetics, acoustics and elastodynamics, and is
equivalent to $f_{0}^{eff}=0$. The coherent potential approximation appears to
be an approximation of the solution to Eq. \eqref{f0eff}\ to first order in
concentration. This leads to two important conclusions. First, the application
of the self-consistent scheme to the quasi-crystalline approximation does not
change the result (\textit{e.g.} we obtain the same effective wavenumber in
both cases, if and only if the effective parameters derived in this paper are
employed), and second, it reduces exactly to the coherent potential
approximation, at least to second order in concentration. The later is in
contrast with the findings of Ref. \cite{Norris2011}, where the wrong
effective parameter was employed in the self-consistent scheme, that is
$\mathsf{m}_{eff}=\mathsf{m}+\phi\left(  \mathsf{m}_{0}-\mathsf{m}\right)  $
(describing the effective mass density of a fluid-particle mixture). Finally,
let us note another important aspect of the self-consistent effective field
method presented in this section. The new self-consistent scheme \eqref{sqca}
includes the influence of the spatial distribution of particles (through the
matrix $\widetilde{{{\mathbf{R}}}}_{eff}$), whereas the coherent potential
approximation does not.

\end{document}